\documentclass[twocolumn,showpacs,preprintnumbers,amsmath,amssymb]{revtex4}
\usepackage{graphicx}% Include figure files
\usepackage{dcolumn}% Align table columns on decimal point
\usepackage{bm}% bold math
\begin{document}
\title{Polarization field in a single-valley strongly-interacting 2D electron system}
\author{V.~T. Dolgopolov and A.~A. Shashkin}
\affiliation{Institute of Solid State Physics, Chernogolovka, Moscow District 142432, Russia}
\begin{abstract}
The magnetic field of complete spin polarization is calculated in
a disorderless single-valley strongly-interacting 2D electron
system. In the metallic region above the Wigner-Mott transition,
non-equilibrium spin states are predicted, which should give rise
to hysteresis in the magnetization.
\end{abstract}
\maketitle

A transition from the high conductivities $\sigma\gg e^2/h$ to the
low conductivities $\sigma\ll e^2/h$ that is observed with
decreasing electron density in disordered 2D electron systems was
first interpreted as the Anderson metal-insulator transition
\cite{ando82}. The interpretation changed after the appearance of
the weak localization theory \cite{gorkov79} and scaling
hypothesis \cite{abrahams79}. According to the latter, there is no
metallic state in disordered infinite 2D electron systems at zero
temperature. As a result, the transition observed in experiment
was referred to as ``apparent'' metal-insulator transition.

The next interpretation change was caused by the finding that the
temperature dependence of the conductivity in low-disordered 2D
electron systems changes sign at a critical density attributed to
the metal-insulator transition point \cite{kravchenko94}. The
observed critical density corresponds to strong electron-electron
interactions, whereas noninteracting or weakly-interacting 2D
systems are considered in the scaling concept. Moreover, the spin
susceptibility and effective mass in the least-disordered 2D
electron systems were found to increase at low densities with a
tendency to diverge at some electron density that is weakly
dependent on disorder \cite{shashkin01,vitkalov01}. This indicates
that the metal-insulator transition observed in a number of 2D
electron systems is driven by interactions \cite{punnoose05},
while the disorder is of minor importance. Additional confirmation
for this statement is provided by the observation of the critical
increase of the effective mass with increasing interactions in the
2D fermion system composed of He$^3$ atoms \cite{casey03}.

There has been published a good deal of experimental and
theoretical work on the metal-insulator transition in two
dimensions and related phenomena (see, e.g., experimental reviews
\cite{kravchenko04,shashkin05,gantmakher08,shashkin10} and
theoretical publications
\cite{spivak04,pankov08,spivak10,terletska11,dobrosavljevic12}).
The key question is whether the metal-insulator transition
observed in low-disordered 2D electron systems is a transition to
the Wigner crystal. The mass divergence when the crystallization
point is approached on the metallic side was already demonstrated
in the paper \cite{dolgopolov02} using Gutzwiller's theory.
Recently, many properties of strongly-interacting 2D and 3D
systems have been explained within the concept of the Wigner-Mott
transition using dynamical mean-field theory
\cite{dobrosavljevic12}.

In this paper, we study the magnetic field of full spin
polarization $B_c$ as a function of electron density $n_s$ for the
Wigner-Mott transition. The magnetic field is assumed to be
parallel to the 2D electron system and act on electron spins only.
We use a lattice model \cite{dolgopolov02} by adapting the results
of the papers \cite{gutzwiller65,brinkman70} to the problem to be
solved. The real 2D electron system is replaced by lattice sites
with density $n_s$. If for the lattice introduced there is an
energy gap between the lowest and next bands, filling every site
with one electron results in a Wigner crystal model. However, in
the ground state, two electrons with anti-parallel spins can be
located, despite repulsion, on each site with some probability
depending on $n_s$. Such sites determine the portion of mobile
electrons and, thus, the transport properties of the electron
system. For the case of noninteracting electrons, the model used
corresponds to the half-filled lowest energy band. The
noninteracting electrons in zero magnetic field occupy all states
up to the Fermi momentum $k_F$, each state being filled with
spin-up and spin-down electrons. In the lattice model, the
probability to find on each site two electrons is equal to $1/4$,
which is the same as the probability to find on each site no
electrons.

\begin{figure}
\scalebox{0.44}{\includegraphics{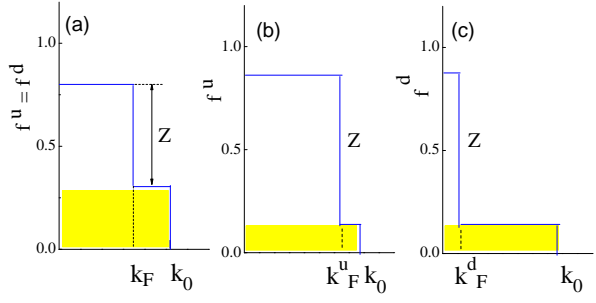}}
\caption{\label{fig1} Probability to occupy a state with momentum
$k$ as a function of momentum in a certain direction: (a) for
spin-unpolarized system; (b) for spin-up electrons; (c) for
spin-down electrons. The shaded region corresponds to the Wigner
crystal.}
\end{figure}

The momentum distribution function for the system with electron
interactions in the metallic state is shown in Fig.~\ref{fig1}(a)
for zero magnetic field. All states are occupied up to the
Brillouin zone boundary $k_0$. A jump of the distribution function
with height $Z<1$ occurs at $k_F$. With decreasing electron
density both $k_0$ and $k_F$ decrease, the ratio $k_F/k_0$ being
constant. As the critical density $n_c$ is approached, the jump
$Z\rightarrow0$ and the distribution function
$f^{u,d}\rightarrow0.5$. The shaded region that spans up to $k_0$
in the figure corresponds to the Wigner crystal. The many-electron
wave function \cite{gutzwiller65} describes the superposition of
the crystal state and the Fermi quasi-particles with an effective
mass $m^*$ that is inversely proportional to the distribution
function jump at the Fermi momentum $k_F$: $m^*/m=1/Z$, where $m$
is the band mass.

The distribution function for electrons with different spins in
magnetic fields is displayed in Fig.~\ref{fig1}(b, c). Both the
quasi-particle weight $Z$ and the Fermi momenta of quasi-particles
with different spins are influenced by magnetic field. Evidently,
the increase of magnetic field leads to $k_F^u\rightarrow k_0$ and
$k_F^d\rightarrow0$. In what follows we find that with increasing
magnetic field the quasi-particle weight $Z$ increases (decreases)
for $n_s\gg n_c$ ($n_s\gtrsim n_c$) until at some degree of spin
polarization, the electron system reaches by jump full spin
polarization in which case every lattice site is filled with one
electron. The presence of the jump of the degree of spin
polarization gives rise to hysteresis with decreasing magnetic
field.

In accordance with the results \cite{gutzwiller65,remark}, the
ground state-averaged Hamiltonian in a parallel magnetic field $B$
is written
\begin{equation} <H>=-\frac{Z}{4D_0}n_s^2(1-p^2)+\frac{\nu\alpha e^2}{\kappa}n_s^{3/2}-\frac{p}{2}n_sg\mu_BB,\label{eq1}\end{equation}
where $\nu$ is the portion of doubly occupied sites ($0<\nu<1/2$),
$e$ is the electron charge, $\kappa$ is the dielectric constant,
$D_0$ is the density of states of the spin-polarized
noninteracting electrons. The degree of spin polarization
$p=(n_u-n_d)/n_s$ is determined by the difference of the densities
of spin-up and spin-down electrons, and the coefficient $\alpha$
is determined by the electron wave function on the lattice site.
The kinetic energy is counted from the band-averaged energy for
noninteracting electrons. The relation between $Z$ and $\nu$
taking account of correlations is as follows
\begin{equation} Z=\frac{2\nu}{1-p^2}\left[(1+p-2\nu)^{1/2}+(1-p-2\nu)^{1/2}\right]^2.\label{eq2}\end{equation}
For $p=0$, Eqs.~(\ref{eq1}, \ref{eq2}) reduce to the known ones in
zero magnetic field \cite{dolgopolov02}.

It is necessary to find a minimum of the expression (\ref{eq1})
over $\nu$, regarding the relation (\ref{eq2}):
\begin{equation} \frac{\partial<H(\nu,p,n_s)>}{\partial\nu}=0,\label{eq3}\end{equation}
which yields a dependence $\nu(p,n_s)$
\begin{equation} 1-4\nu+\frac{(1-2\nu)(1-4\nu)-p^2}{[(1-2\nu)^2-p^2]^{1/2}}=2\left(\frac{n_c}{n_s}\right)^{1/2},\label{eq4}\end{equation}
where $n_c=(\alpha e^2D_0/2\kappa)^2$.

Note that for $p=0$, the ground state energy is a minimum if
\begin{equation} \nu=\frac{1}{4}\left[1-\left(\frac{n_c}{n_s}\right)^{1/2}\right].\label{eq5}\end{equation}
At $n_s\rightarrow n_c$, both $\nu$ and $Z$ tend to zero, i.e.,
every lattice site is filled with one electron, and the
quasi-particle mass diverges
\begin{equation} \frac{m^*}{m}=\frac{n_s}{n_s-n_c}.\label{eq6}\end{equation}
In the opposite limiting case of $n_s\rightarrow\infty$, the
values $\nu$ and $Z$ are $\nu=1/4$ and $Z=1$, and the
quasi-particle mass is equal to $m^*=m$.

\begin{figure}
\scalebox{0.44}{\includegraphics{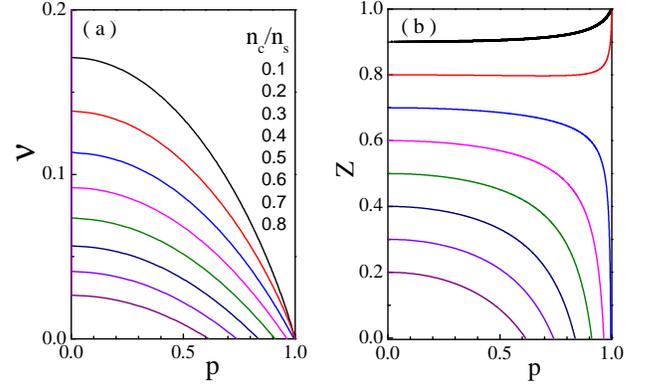}}
\caption{\label{fig2} (a) Portion of doubly occupied sites and (b)
the quasi-particle weight versus spin polarization at different
values of the interaction parameter.}
\end{figure}

The dependence $\nu(p)$ is shown in Fig.~\ref{fig2}(a) for
different values of the interaction parameter $n_c/n_s$. If
$n_s>4n_c$, the value $\nu$ zeroes only at $p=1$, whereas at
$n_s<4n_c$ the equality $\nu=0$ is the case over a whole range of
$p$. We note that this fact is important for determining the
dependence $B_c(n_s)$, and the density $n_s=4n_c$ manifests itself
as ``apparent'' critical density. The corresponding dependence
$Z(p)$ is represented in Fig.~\ref{fig2}(b). The value $Z$
increases (decreases) with spin polarization at $n_s>4n_c$
($n_s<4n_c$), attaining the limiting value $Z=1$ ($Z=0$).
Interestingly, the spin states near $p=1$ turn out to be
non-equilibrium states. If the dependence $<H(p)>$ is two minima
separated by a maximum, the electron system at $T=0$ gets to the
deeper minimum in a magnetic field where the maximum disappears,
which causes hysteresis.

We shall start by considering the high-density region $n_s>16n_c$.
In the spirit of the paper \cite{fleury10}, one needs to find a
minimum of the Hamiltonian (\ref{eq1}, \ref{eq2}) over $p$:
\begin{equation} \frac{\partial<H(\nu,p,n_s)>}{\partial p}=0,\label{eq7}\end{equation}
which yields
\begin{equation} \frac{2\nu p}{[(1-2\nu)^2-p^2]^{1/2}}=\frac{g\mu_BBD_0}{n_s}.\label{eq8}\end{equation}
By solving Eqs.~(\ref{eq4}, \ref{eq8}) with $p\rightarrow1$, we
get the polarization field
\begin{equation} B_c=\frac{n_s}{g\mu_BD_0}\left[1-2\left(\frac{n_c}{n_s}\right)^{1/2}\right].\label{eq9}\end{equation}
The dependence obtained is the same as the one suggested earlier
\cite{shashkin01} and is similar to $B_c(n_s)$ observed in
experiment. Obviously, Eq.~(\ref{eq9}) cannot describe the
equilibrium value $B_c$ down to $n_s=4n_c$, because the
Wigner-Mott transition in zero magnetic field occurs at the lower
density $n_s=n_c$.

\begin{figure}
\scalebox{0.42}{\includegraphics{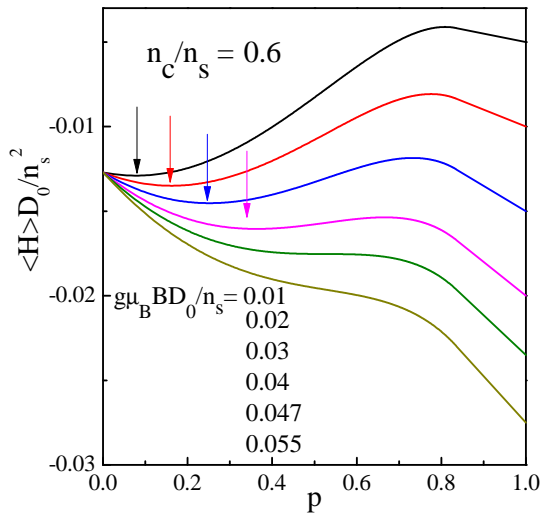}}
\caption{\label{fig3} The ground state energy as a function of
spin polarization at fixed $n_c/n_s=0.6$ for different magnetic
fields. The minimum is marked by arrows.}
\end{figure}

\begin{figure}
\scalebox{0.44}{\includegraphics{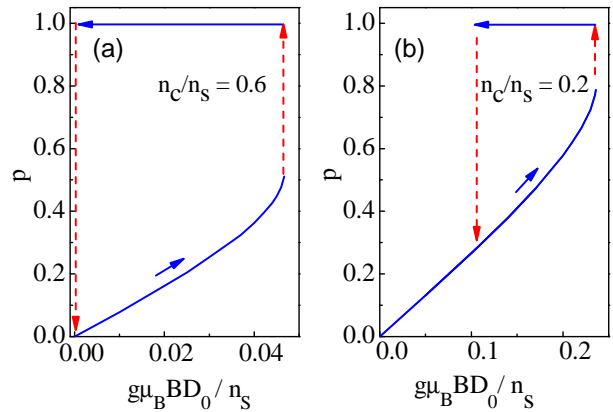}}
\caption{\label{fig4} Hysteresis of the degree of spin
polarization with changing magnetic field at fixed (a)
$n_c/n_s=0.6$ and (b) $n_c/n_s=0.2$.}
\end{figure}

We solve the problem numerically at electron densities near
$4n_c$. There, the system behavior turns out to be hysteretic.
Using the dependence $\nu(p)$ of Fig.~\ref{fig2}, we calculate the
dependence $<H(p)>$ for a fixed value of the interaction parameter
$n_c/n_s$ at different magnetic fields, as shown in
Fig.~\ref{fig3} for $n_c/n_s=0.6$. In weak magnetic fields, the
energy has a minimum corresponding to the realized degree of spin
polarization. As the magnetic field is increased, the minimum
turns into an inflection and disappears entirely. The inflection
determines the upper polarization field at which the system state
jumps to $p=1$, $\nu=0$. The degree of spin polarization first
increases linearly with magnetic field, then turns up, and
finally, the electron system reaches complete spin polarization by
jump, as shown in Fig.~\ref{fig4}. With decreasing magnetic field
there arises hysteresis. The electron system stays in the $p=1$
state until the maximum at $p<1$ in the dependence $<H(p)>$
(Fig.~\ref{fig3}) disappears at the lower polarization field at
which the system state jumps out of $p=1$ to close the hysteresis
loop (Fig.~\ref{fig4}). It is clear that the lower polarization
field is given by Eq.~(\ref{eq9}) at $n_s>4n_c$ and is equal to
zero at $n_s<4n_c$.

\begin{figure}[b]
\scalebox{0.4}{\includegraphics{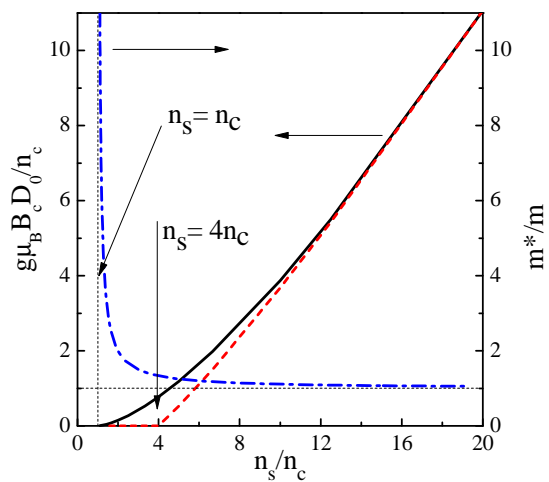}}
\caption{\label{fig5} Polarization field as a function of electron
density with increasing (solid black line) and decreasing (dashed
red line) magnetic field. Also shown by the dash-dotted blue line
is the effective mass in $B=0$ versus electron density.}
\end{figure}

The dependence of the polarization field on electron density with
increasing and decreasing magnetic field is displayed in
Fig.~\ref{fig5} along with the effective mass as a function of
electron density in zero magnetic field. Based on linear
extrapolation of the high-density values, the field $B_c$ tends to
zero at $n_s\approx4n_c$, which is well above the density where
the effective mass in zero magnetic field diverges. Note that very
low temperatures are needed to observe the hysteresis in $B_c$
because the maximum in Fig.~\ref{fig3} is smeared due to thermal
fluctuations already at $T\sim0.1$~K.

In accordance with the solution, one might conclude that the
magnetic field promotes crystallization at $n_s<4n_c$, whereas at
higher densities, the system state in magnetic fields is the
electron liquid. As a matter of fact, the complete filling of the
first Brillouin zone in the model used means crystallization only
if the energy gap is present for the model lattice. In turn, the
energy gap is determined by electron-electron interactions on
different lattice sites so that in the frames of the model
\cite{dolgopolov02,gutzwiller65,brinkman70}, one cannot say
whether or not the energy gap is present. Nevertheless, one can
expect that the prediction is qualitatively correct: the gap in
magnetic fields should survive at electron densities above $n_c$
and disappear at yet higher densities. Within the hysteresis
uncertainty, the low-density part of $B_c(n_s)$ separates the
correlated electron liquid and crystal and corresponds to the
metal-insulator transition. The last fact is confirmed by
comparison of the experimental results of Ref.~\cite{shashkin01}
and Refs.~\cite{dolgopolov92,shashkin02}.

Another prediction of our calculations is that the effective mass
in a single-valley electron system should decrease with magnetic
field and reach the noninteracting electron mass at the onset of
complete spin polarization if the system stays in the metallic
state. The same behavior of the mass should be the case for full
isospin polarization in a two-valley spin-polarized electron
system. According to the statements of the papers
\cite{padmanabhan08,gokmen10}, the effect has been observed
recently in the 2D electron system in AlAs quantum wells.

The magnetic field of full spin polarization as a function of
electron density in the single-valley 2D electron system of
GaAs/AlGaAs heterostructures was studied in Ref.~\cite{zhu03}.
Because of large thickness of the 2D system, the dependence
$B_c(n_s)$ is strongly distorted by orbital effects. Nevertheless,
its slope for low $B_c$ is in agreement with the slope expected
from Eq.~(\ref{eq9}), and $B_c(n_s)$ extrapolates to zero at a
finite density \cite{shashkin05}.

The most complete and detailed experimental information has been
obtained for two-valley 2D electron systems. However, our
calculations are not relevant for such systems, and comparison
with those experiments is not justified.

It is interesting to compare our results with alternative
calculations. The behavior of the effective mass near the critical
density of Eq.~(\ref{eq6}) has been reproduced in
Ref.~\cite{dobrosavljevic12}. There, the dependence $B_c(n_s)$ has
also been obtained which, unlike our results, does not reveal
hysteresis.

It is worth noting that the polarization field $B_c$ versus $n_s$
for a two-valley 2D electron system has been calculated using
quantum Monte-Carlo simulations \cite{fleury10}. In the clean
limit ($k_Fl\gg1$, where $l$ is the mean free path), the
divergence of the mass and the critical $B_c(n_s)$ are not found.
However, it is strange that the degree of spin polarization in the
paper \cite{fleury10} is proportional to magnetic field,
regardless of disorder and interaction strength.

In summary, we have calculated the dependence of the magnetic
field of complete spin polarization on electron density for the
Wigner-Mott transition in a single-valley 2D electron system. The
following predictions of the model used have not yet been
confirmed in experiment. In the metallic region above the
Wigner-Mott transition, non-equilibrium spin states are expected,
which should lead to a hysteretic behavior of the magnetization.
Linear extrapolation to zero of the high-density values of the
polarization field yields a critical density that exceeds the
Wigner-Mott transition point by the factor of about four.

We gratefully acknowledge discussions with I.~S. Burmistrov, V.
Dobrosavljevic, and S.~V. Kravchenko. This work was supported by
RFBR, RAS, and the Russian Ministry of Sciences.

\end{document}